\documentclass[]{spie}  %>>> use for US letter paper
\usepackage{amsmath,amsfonts,amssymb}
\usepackage{graphicx}
\usepackage[colorlinks=true, allcolors=blue]{hyperref}
\usepackage{float}

\title{Fast, low noise CCD systems for future strategic x-ray missions}

\author[a,b]{Haley R. Stueber}
\author[a,b]{Abigail Y. Pan}
\author[a]{Tanmoy Chattopadhyay}
\author[a,b,c]{Steven W. Allen}
\author[d]{Marshall W. Bautz}
\author[e]{Kevan Donlon}
\author[d]{Catherine E. Grant}
\author[a]{Sven Herrmann}
\author[d]{Beverly J. LaMarr}
\author[d]{Andrew Malonis}
\author[d]{Eric D. Miller}
\author[a,c]{R. Glenn Morris}
\author[a]{Peter Orel}
\author[a]{Artem Poliszczuk}
\author[d]{Gregory Y. Prigozhin}
\author[a,b]{Daniel R. Wilkins}
\affil[a]{Kavli Institute for Particle Astrophysics and Cosmology, Stanford University, 452 Lomita Mall, Stanford, CA 94305, USA}
\affil[b]{Department of Physics, Stanford University, 382 Via Pueblo Mall, Stanford CA 94305, USA}
\affil[c]{SLAC National Accelerator Laboratory, 2575 Sand Hill Road, Menlo Park, CA 94025, USA}
\affil[d]{MIT Kavli Institute for Astrophysics and Space Research, Massachusets Institute of Technology, 70 Vassar St, Cambridge, MA 02139, USA}
\affil[e]{MIT Lincoln Laboratory, 244 Wood St building 1324, Lexington, MA 02421, USA}

\authorinfo{Further author information: Send correspondence to H.R. Stueber\\H.R. Stueber: E-mail: hstueber@stanford.edu, Telephone: 1 (920)-252-5189}

\begin{document} 
\maketitle

\begin{abstract}
Future strategic X-ray missions, such as the Advanced X-ray Imaging Satellite (AXIS) and those targeted by the Great Observatories Maturation Program (GOMaP), require fast, low-noise X-ray imaging spectrometers. To achieve the speed and noise capabilities required by such programs, the X-ray Astronomy and Observational Cosmology (XOC) Group at Stanford, in collaboration with the MIT Kavli Institute (MKI) and MIT Lincoln Laboratory (MIT-LL), is developing readout systems that leverage the high speed, low noise, and low power consumption of application-specific integrated circuit (ASIC) devices. Here, we report the energy resolution and noise performance achieved using MIT-LL AXIS prototype charge-coupled device (CCD) detectors in conjunction with Stanford-developed Multi-Channel Readout Chip (MCRC) ASICs. Additionally, we present a new sampling method for simultaneous optimization of the output gate (OG), reset gate (RG), and reset drain (RD) biases which, in combination with new integrated fast summing well (SW) and RG clock operation modes, enables the data rates required of future X-ray telescopes.  
\end{abstract}


\section{INTRODUCTION}
\label{sec:intro}  

Next generation X-ray observatories such as the proposed Advanced X-ray Imaging Satellite (AXIS)\cite{chrisSPIE2023} and those targeted by the Great Observatories Maturation Program (GOMaP) will enable transformative studies of the low luminosity and high redshift X-ray universe, providing insight into the formation and evolution of black holes, the role of gas flow in galaxy formation, and explosive transient phenomena. To accomplish this, these observatories will need to be equipped with sensitive, high spatial resolution, wide field-of-view imaging detectors. Mitigating pile-up from bright sources and suppressing the impact of the particle background in observations of faint, diffuse sources will be critical, and will require that these imagers achieve high frame rates while maintaining low noise and excellent soft energy response. Active Pixel Sensors (APS) such as Hybrid CMOS detectors (HCDs) \cite{HCMOS07, HCMOS17} and Depleted Field Effect Transistor (DEPFET) detectors \cite{DEPFET20} such as those being developed for the Athena Wide Field Imager (WFI) \cite{athenaSPIE2017} have been shown to deliver excellent performance in some of these areas. However, contemporary HCDs are limited by relatively high readout noise \cite{chattopadhyay18_HCDoverview}, while DEPFET detectors tend to have relatively large pixel sizes, making them less suitable candidates for missions targeting high spatial resolution over large collecting areas. Current state-of-the-art X-ray charge-coupled device (CCD) detectors, on the other hand, with their smaller pixels and low noise levels, can deliver the performance requirements for missions like AXIS, except for the frame rates.

The X-ray Astronomy and Observational Cosmology (XOC) group at Stanford in collaboration with colleagues at the Massachusetts Institute of Technology (MIT) and MIT Lincoln Laboratory (MIT-LL) is addressing the technology gap of fast, low-noise X-ray imaging CCD's. In this manuscript, we focus on four distinct approaches being taken to achieve the bandwidth, frame rates, noise, and power requirements of next generation astronomical X-ray missions:

\begin{itemize}
\item Developing a fast and low noise CCD output stage with multiple output nodes, needed to read large format detectors at high speeds.
\item Engineering application specific integrated circuit (ASIC) readout electronics that enable parallel readout and provide comparable or better speed and noise performance to discrete electronics solutions at a fraction of the footprint and power consumption. 
\item Optimizing noise and speed of the detector clocks and biases and their driving circuits.
\item Advancing analysis techniques to determine optimal biases of the detector sense node and of the ASIC readout electronics.
\end{itemize}

% The next generation of astronomical X-ray missions will aim to expand on the success of Chandra by combining high angular resolution with large collecting area and a wide field of view, to probe deeper into the low luminosity and high redshift X-ray universe

\section{FAST, LOW NOISE X-RAY CCDs}
\label{sec:CCDs}

%To enable the full discovery potential of future strategic X-ray satellite missions such as AXIS, large X-ray imaging detectors are needed. These detectors must be fast to avoid pileup, and low noise to allow for faint source detection. 
The large area detectors of next generation observatories will require frame rates at least an order of magnitude faster than those of legacy observatories such as Chandra. To this end, the Stanford XOC group, in collaboration with the MKI and MIT-LL, are developing fast, low-noise X-ray detectors and associated readout technology. In particular, we are currently characterizing small-scale AXIS prototype MIT-LL CCID-93 detectors and full-scale prototype CCID-100 devices. 
%These devices benefit from single-level polysilicon process fabrication technology, which enables low power operation of the CCD's and offers low capacitance for achieving faster transfer speeds. They also feature fast, high responsitivity multi-stage detector outputs. 
To read out these detectors and optimize their speed and noise performances, our team is also developing fast, low power, low noise, small footprint application-specific integrated circuit (ASIC) chips. 
%The multi-channel readout chip (MCRC) ASIC enables fast, parallel multi-channel readout, saving power and board space compared to discrete solutions. 
In this section, we will provide an overview of the detectors and our readout solution.

\subsection{MIT-LL CCID-93 Detectors}
\label{sec:CCID93}

\begin{figure*}[ht!]
   \begin{center}
   \begin{tabular}{c}
   \includegraphics[width=0.5\textwidth]{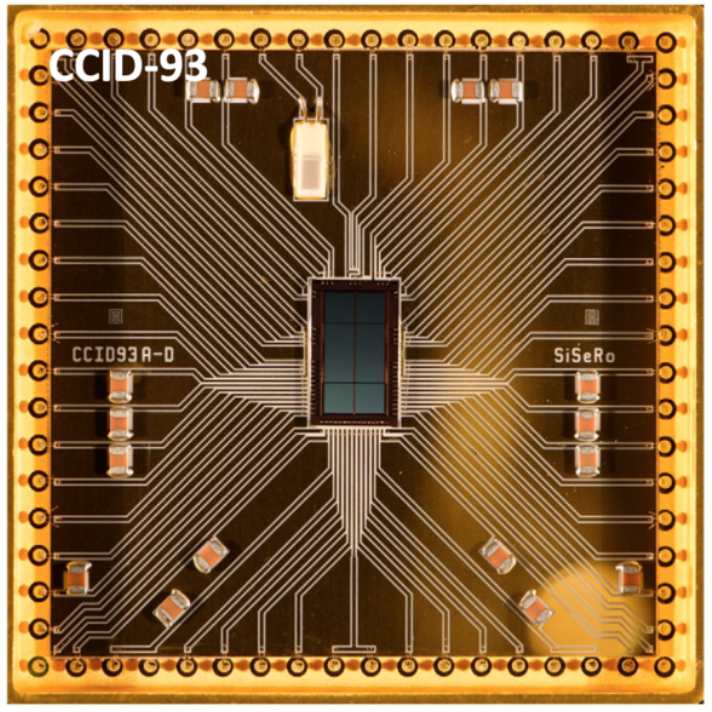}
   \includegraphics[width=0.335\textwidth]{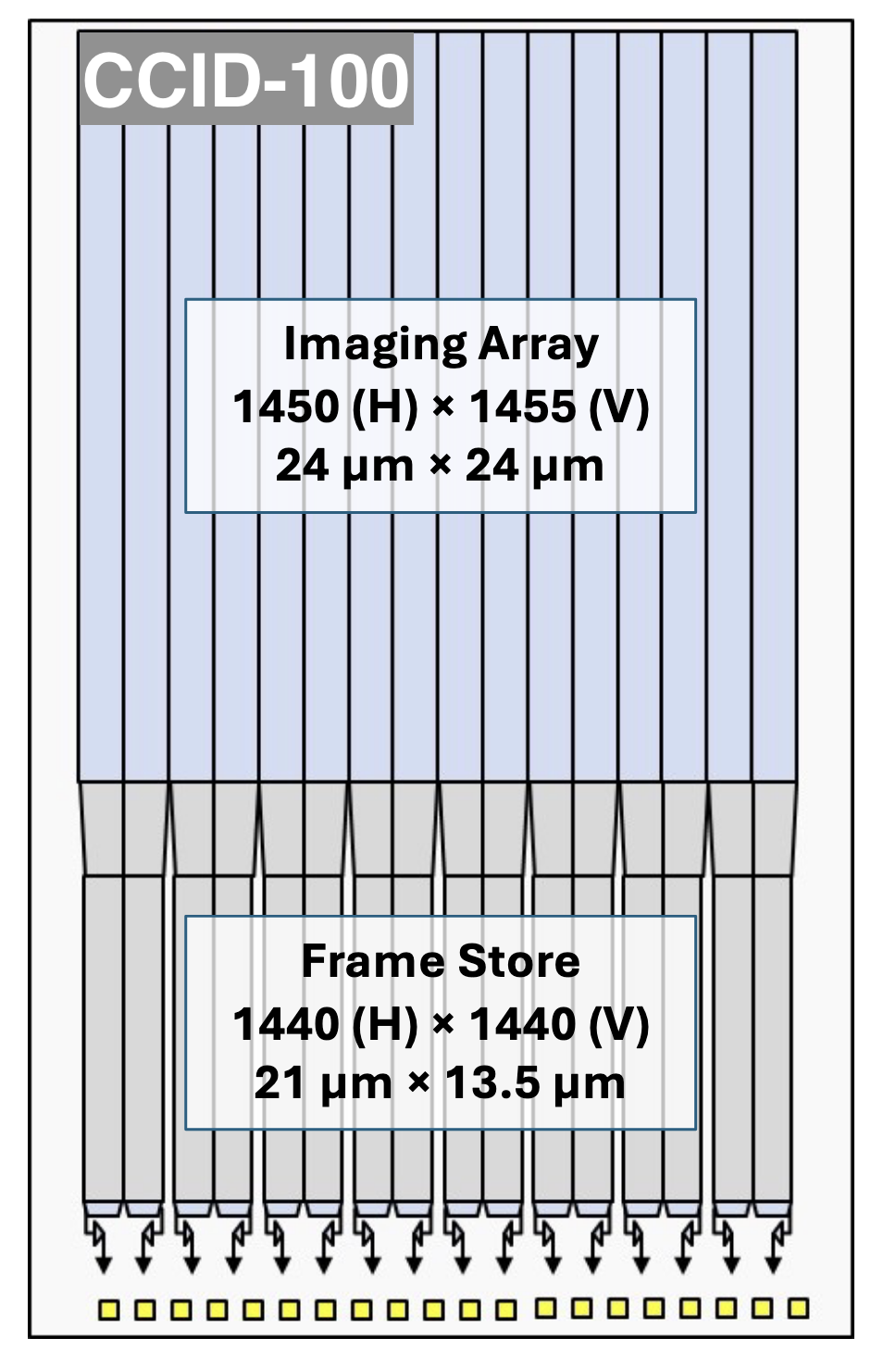}
   \end{tabular}
   \end{center}
   \caption[example] 
   { \label{fig:CCID93_MCRC} 
{\it Left:} Image of an MIT-LL CCID-93 X-ray detector.\cite{bautz2024} The detector is 512 x 512 8 $\rm \mu m$ pixels in the imaging area, with an identically sized frame store. This is a scaled-down prototype of the AXIS detectors. {\it Right:} Schematic diagram of the MIT-LL CCID-100 layout.\cite{bautz2024}. With $ \rm 1440\,x\,1440$ $\rm 24\,\mu m$ imaging pixels and 16 output channels, this is a full-scale AXIS prototype detector. }
\end{figure*}

\begin{figure*}[ht!]
   \begin{center}
   \begin{tabular}{c}
   \includegraphics[width=0.7\textwidth]{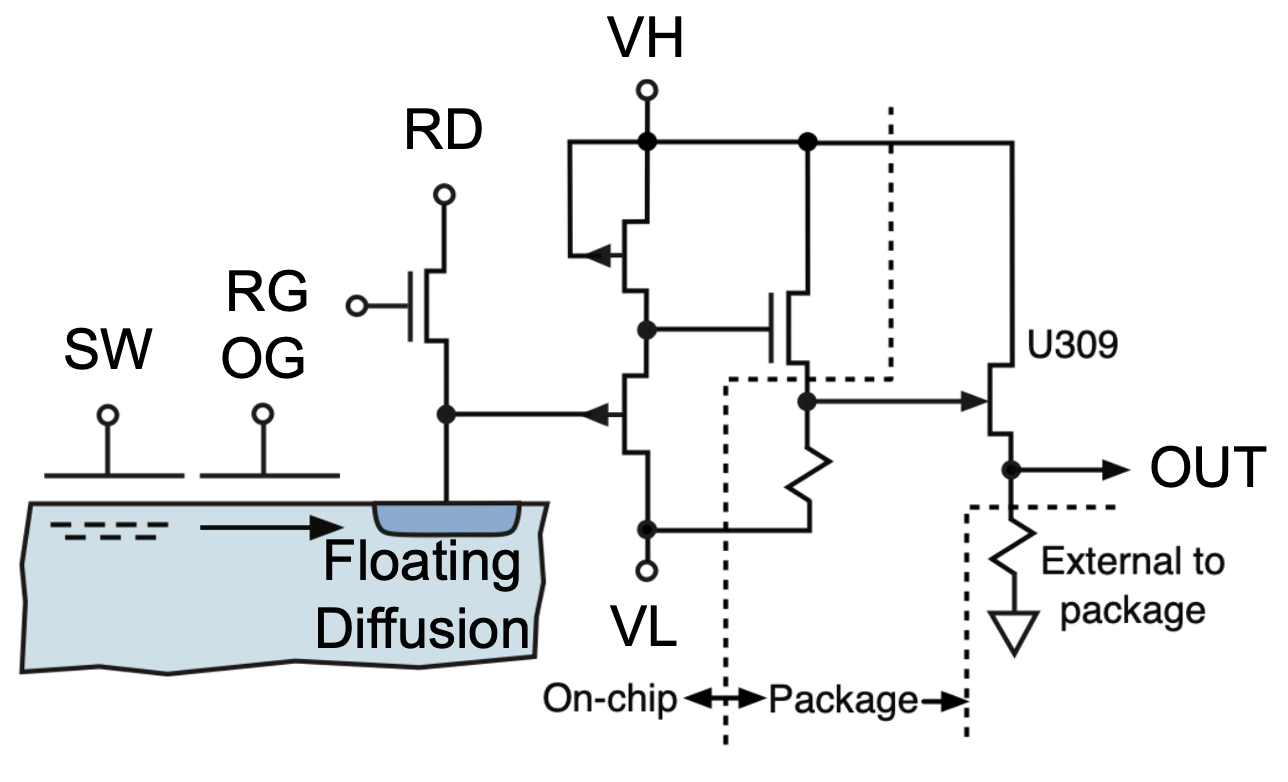}
   \end{tabular}
   \end{center}
   \caption[example] 
   { \label{fig:output_stage} 
Output stage of the CCID-93. Charge from the CCD is transferred to the Floating Diffusion implant, followed by a two-stage output featuring a source follower pJFET transistor first stage and an nMOSFET transistor second stage\cite{tanmoyJATIS22}. The Reset Gate (RG), Output Gate (OG), and Reset Drain (RD) nodes are labeled. 
}
\end{figure*}

The MIT-LL CCID-93 is a small format, front-illuminated detector with 8 micron square pixels comprising a $\rm 512\,x\,512$ pixel imaging area and a frame store of the same size. An image of the CCID-93 detector is shown on the left side of Figure \ref{fig:CCID93_MCRC}\cite{bautz2024}. Detailed descriptions of the design and fabrication of the CCID-93 detectors can be found in [\citenum{bautz18}], [\citenum{bautz19}], [\citenum{bautz20}], and [\citenum{gregory2020SPIE}]. In brief, the CCID-93 is a three-phase device in its horizontal and vertical gate structure, where all gates are fabricated with a single layer of polysilicon. Plasma-etched gaps physically isolate the gates. The small size of the gaps in comparison to those in more traditional multi-polysilicon structures requires smaller clocking amplitudes for charge transfer, which consumes less power. This architecture is a critical component in achieving the fast, low noise, low power detector performance required of missions like AXIS.

Another critical component is the output stage of the CCID-93. These detectors have two variants of on-chip output amplifier stages, featuring a source follower p-channel junction field effect transistor (pJFET) that performs voltage readout, as well as the Single electron Sensitive Read Out (SiSeRO) output that performs drain current readout. While not the focus of this paper, the SiSeRO architecture and performance results are reported in [\citenum{sisero2021}], [\citenum{sisero2022}], [\citenum{sisero2023}], [\citenum{tanmoyspie2024}], and [\citenum{tanmoySPIE2025}]. The pJFET output stage, which will be the focus of this paper, is pictured in Figure \ref{fig:output_stage}\cite{tanmoyJATIS22}. Charge from the CCD is transferred to the Floating Diffusion implant, which is followed by a two-stage output featuring a fast, high conversion gain source follower pJFET first stage and an nMOSFET second stage with large bandwidth. The CCID-93 is a first generation proof-of-concept device that successfully meets the speed and noise requirements of AXIS. The next step is the large format CCID-100 detector, which is a full-scale prototype for the AXIS focal plane\cite{millerAXIS2023} and, potentially, other future missions.

\subsection{MIT-LL CCID-100 Detectors}
\label{sec:CCID100}

The initial test version of the MIT-LL CCID-100 device is a front-illuminated detector. It has an imaging area of $\rm 1440\,x\,1440$ pixels and a frame store of the same size, where the pixel pitch is $\rm 24\,\mu m$. While the output stage of the CCID-100 is the same as that of the CCID-93, each CCID-100 has 16 parallel output channels to increase the data rate of its larger imaging area. A schematic diagram of the CCID-100 is shown on the right side of Figure \ref{fig:CCID93_MCRC}. To achieve mission baseline performance, the detector requires fast, parallel readout of all 16 channels concurrently.

\section{Readout Electronics}
\label{sec:MCRC}

\begin{figure*}[ht!]
   \begin{center}
   \begin{tabular}{c}
   \includegraphics[width=0.6\textwidth]{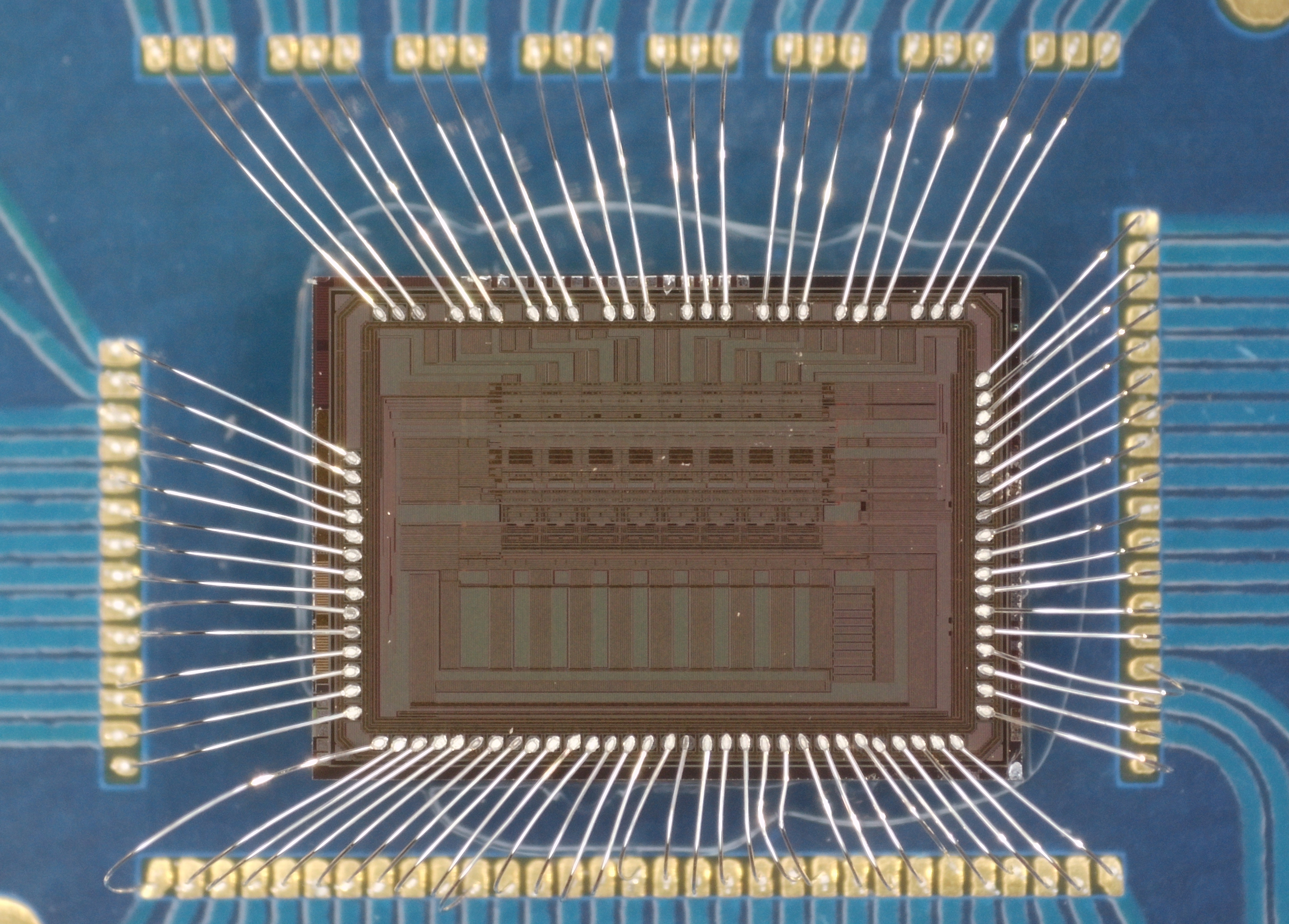}
   \includegraphics[width=0.35\textwidth]{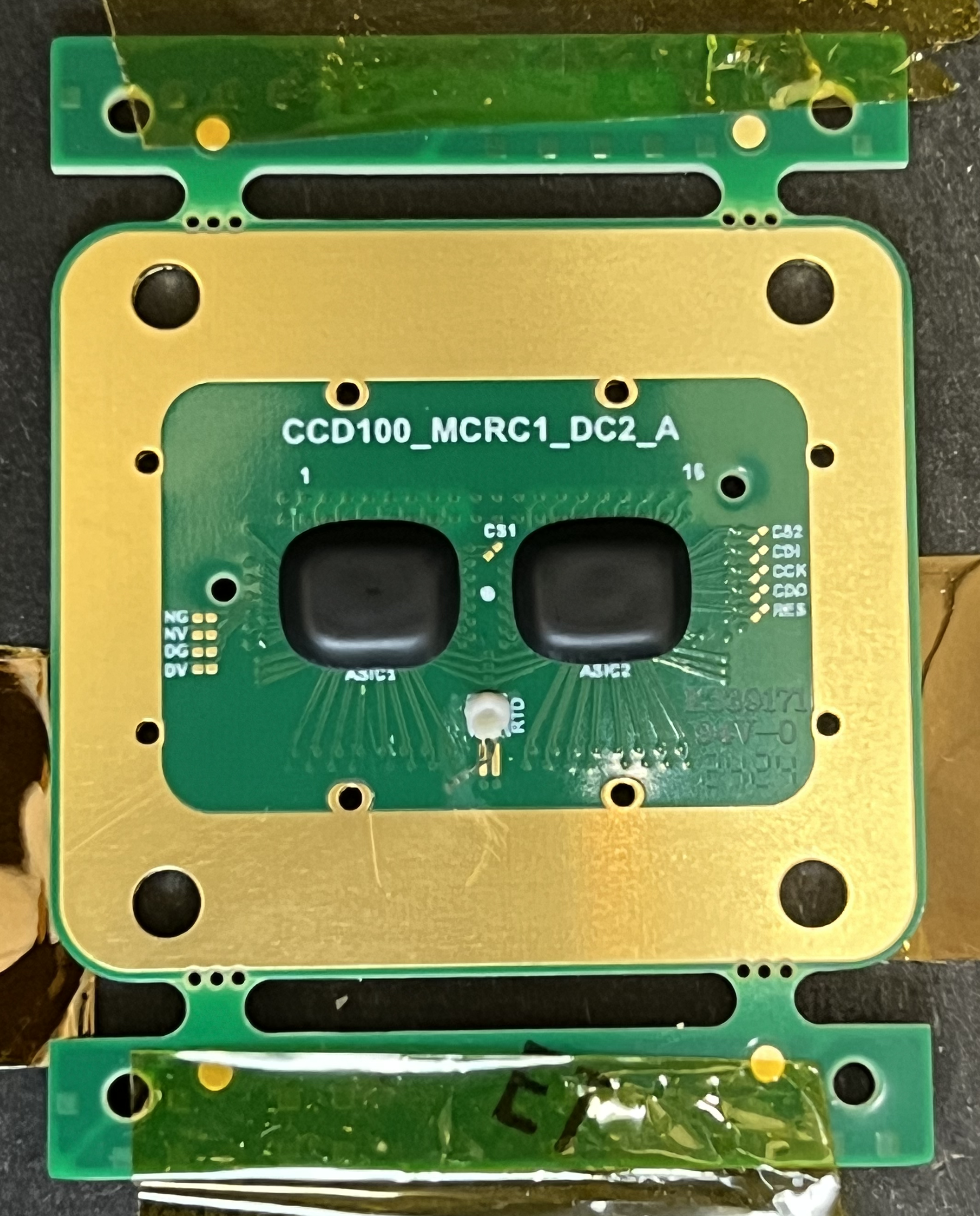}
   \end{tabular}
   \end{center}
   \caption[example] 
   { \label{fig:MCRC} 
{\it Left:} The Stanford Multi-Channel Readout Chip (MCRC) Application-Specific Integrated Circuit (ASIC). The MCRC ASIC has 8 channels for fast readout with minimal noise, power consumption, and physical footprint\cite{porelMCRCspie2022, porelMCRCspie2024}. {\it Right:} Dual-mounted MCRC 8-channel readout chips, which enable 5-20 frames per second readout of 16-channel CCID-100 detectors.
}
\end{figure*}

\noindent The Stanford-developed Multi-Channel Readout Chip (MCRC)\cite{herrmann20_mcrc,porelMCRCspie2022} is an analog ASIC used for fast, low noise readout of MIT-LL CCDs. An image of a fabricated MCRC-V1 chip is pictured on the left side of Figure \ref{fig:MCRC}. It has physical dimensions of $\rm 4160\,\mu m$ x $\rm 2900\,\mu m$ and features 8 analog readout channels that operate in parallel. Full details of the architecture and fabrication of the MCRC ASIC can be found in [\citenum{porelMCRCspie2022}] and [\citenum{herrmann20_mcrc}] with the latest performance reported in [\citenum{porelMCRCspie2024}]. In summary, each of the analog channels of the MCRC consists of an input stage that provides a suitable interface (biasing and impedance) for the output stage of the detector. The user can select between voltage or current inputs for either pJFET or SiSeRO based detector outputs. The input stage is followed by a preamplifier that converts signals from single-ended to differential values and provides user-selectable gain settings of 8 or 16 V/V, respectively. The preamplifier output is buffered by a unit-gain, fully differential output buffer, designed to drive a 100 $\Omega$ transmission line up to 1 meter in length. The MCRC signal waveform is sampled by an analog-to-digital (ADC) converter, and image data is extracted via a digital pulse processing (DPP) algorithm. Both the ADC and the DPP are hosted by an Archon controller\footnote{http://www.sta-inc.net/archon/}. A digital serial peripheral interface (SPI) is used to program the ASIC settings, including control of the internal switch logic and the digital-to-analog converters (DACs) that provide biasing to the internal analog circuitry. Integrating multiple functions, the ASIC streamlines CCD board design by reducing the physical footprint and the number of discrete components. It has a large bandwidth and matches the speed and noise performance of our best discrete readout solutions for a fraction of the power consumption\cite{porelMCRCspie2022, porelMCRCspie2024}. The unique features of the MCRC ASIC give it the capability to deliver and exceed baseline requirements in both speed and noise for future large-format X-ray imagers, such as the high-speed camera on AXIS. Two 8-channel ASIC chips will be used to read out the 16-channel AXIS CCID-100 detector (see the dual-ASIC board pictured on the right side of Figure \ref{fig:MCRC}) at frame rates of up to 20 frames/s.

\subsection{Onboard Fast Summing Well and Reset Gate Clocks}
\label{sec:SWRG}

\begin{figure*}[ht!]
   \begin{center}
   \begin{tabular}{c}
   \includegraphics[width=0.95\textwidth]{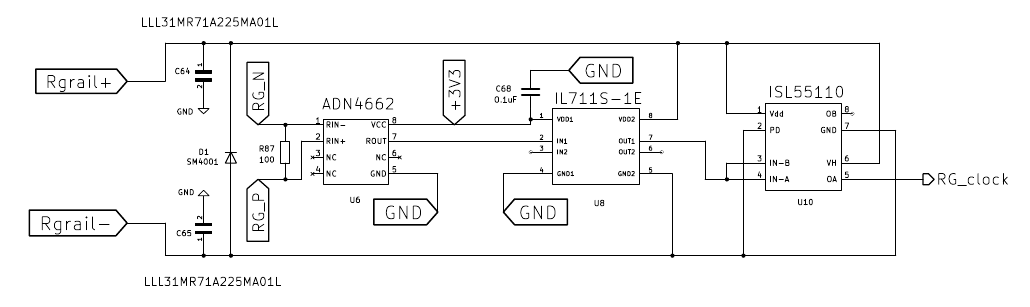}
   \end{tabular}
   \end{center}
   \caption[example] 
   { \label{fig:RG_circuit} 
Circuit schematic for the integrated fast Reset Gate (RG) clock driver. Differential input signals are converted to single-ended, and the negative input is referenced to ground.}
\end{figure*}

\noindent The Reset Gate (RG) and Summing Well (SW) clocks provided by the external Archon controller %which is used to supply biases and clocks to the CCD's as well as digitize and read out their outputs. 
initially limited the readout speed of our prototype CCID-93 test devices to 4\,MHz\cite{tanmoyJATIS22}. To enable faster serial transfer speeds and reduce the noise levels at lower speeds, the SW and RG clock signals have been transitioned to a locally buffered scheme. The onboard circuit for the RG clock signal is pictured in Figure \ref{fig:RG_circuit}; an identical circuit is used to source the SW clock. The onboard circuit converts the clock signals from differential to single-ended form. The single-ended clocks are opto-isolated to enable ground shifting and buffered to provide adjustable amplitudes and offsets for ideal point-of-load driving of the CCD clock inputs.

\begin{figure*}[ht!]
   \begin{center}
   \begin{tabular}{c}
   \includegraphics[width=0.8\textwidth]{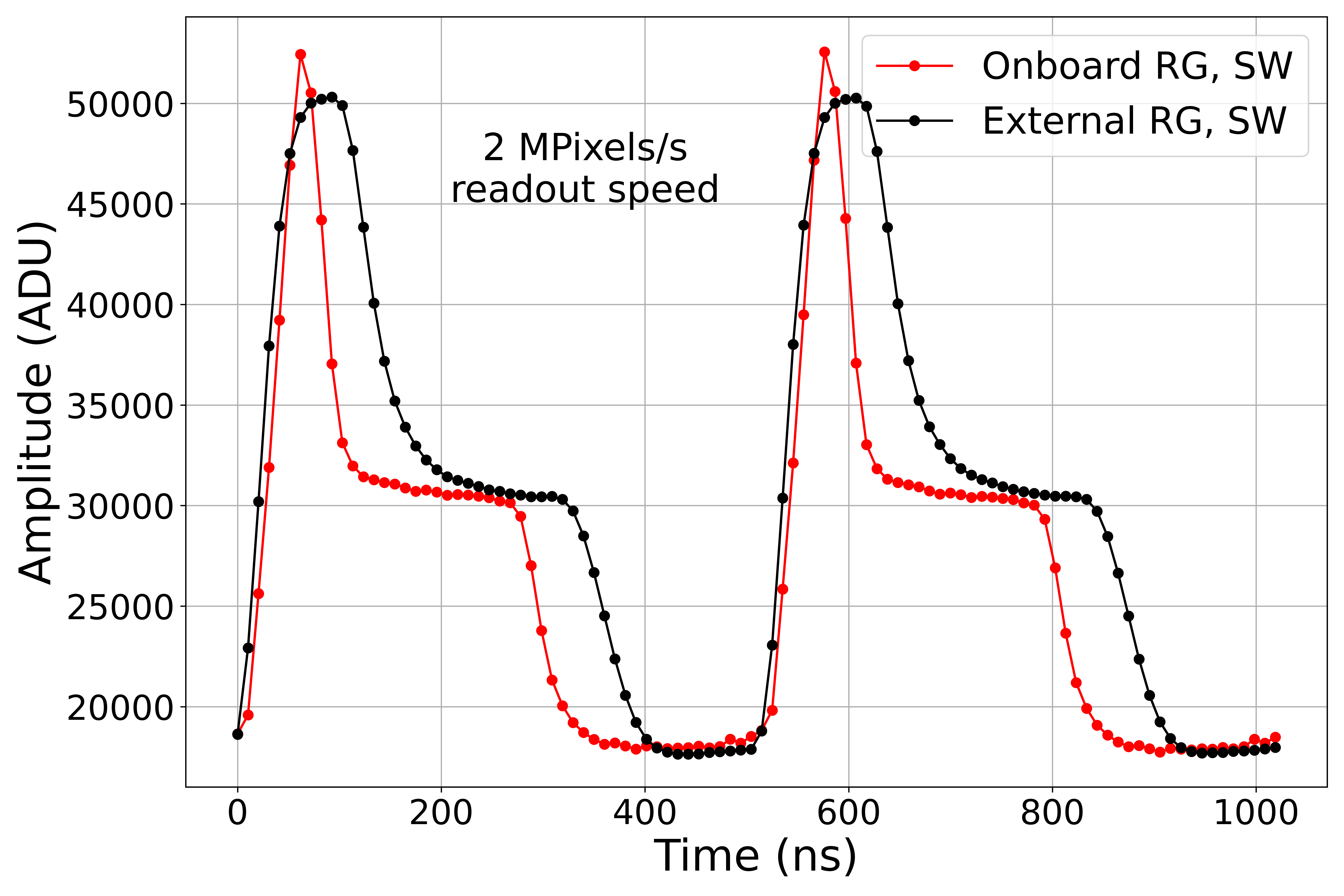}
   \end{tabular}
   \end{center}
   \caption[example] 
   { \label{fig:WF} 
2 MPixels/s waveforms using the external RG and SW drivers (black) and the onboard RG and SW drivers (red). We see $\sim20\%$ recovery in samples in the baseline and the signal region of the waveform using the fast onboard drivers.}
\end{figure*}

The high speed and high current driving capability of the new onboard clock circuitry, combined with a lack of termination, initially introduced ringing in the clock signals. This was mitigated using a "snubber" circuit. A snubber is a form of termination, which in this case is composed of an RC circuit that attenuates the reflected signal and dampens the ringing without significantly sacrificing clock speed. The RC values were determined based on the measured parasitic parallel capacitance and the series inductance, which was derived from the oscillations measured in the clocks. An image comparing the waveform of the CCD with the externally driven SW and RG clocks against the waveform of the CCD with the optimized onboard SW and RG clocks is given in Figure \ref{fig:WF}. We see a $\rm \sim20\%$ recovery of the baseline and signal samples using the faster onboard clocks. These additional samples enable faster serial transfer speeds of up to 5\,MPixels/s. Figure \ref{fig:2MHz_5MHz_WF} presents a 5\,MPixels/s waveform from the CCID-93 detector using the onboard SW and RG clock drivers, compared to the 2 MPixel/s waveform.

Alongside enhancing the frame rate capabilities of the detectors, the onboard clocks also improve noise performance at lower speeds. The close proximity of the onboard drivers to the output circuitry %locality of the onboard clock
reduces electromagnetic interference, improving the noise from the externally driven solution. A summary of the read noise values achieved with the onboard clocks at different readout speeds is given in Table \ref{tab:SWRG_noise}.

\section{OG, RG, RD Bias Scanning and Optimization}
\label{sec:param_scan}

% \subsection{Methods and Implementation}
% \label{sec:implementation}

\begin{figure*}[ht!]
   \begin{center}
   \begin{tabular}{c}
   \includegraphics[width=0.95\textwidth]{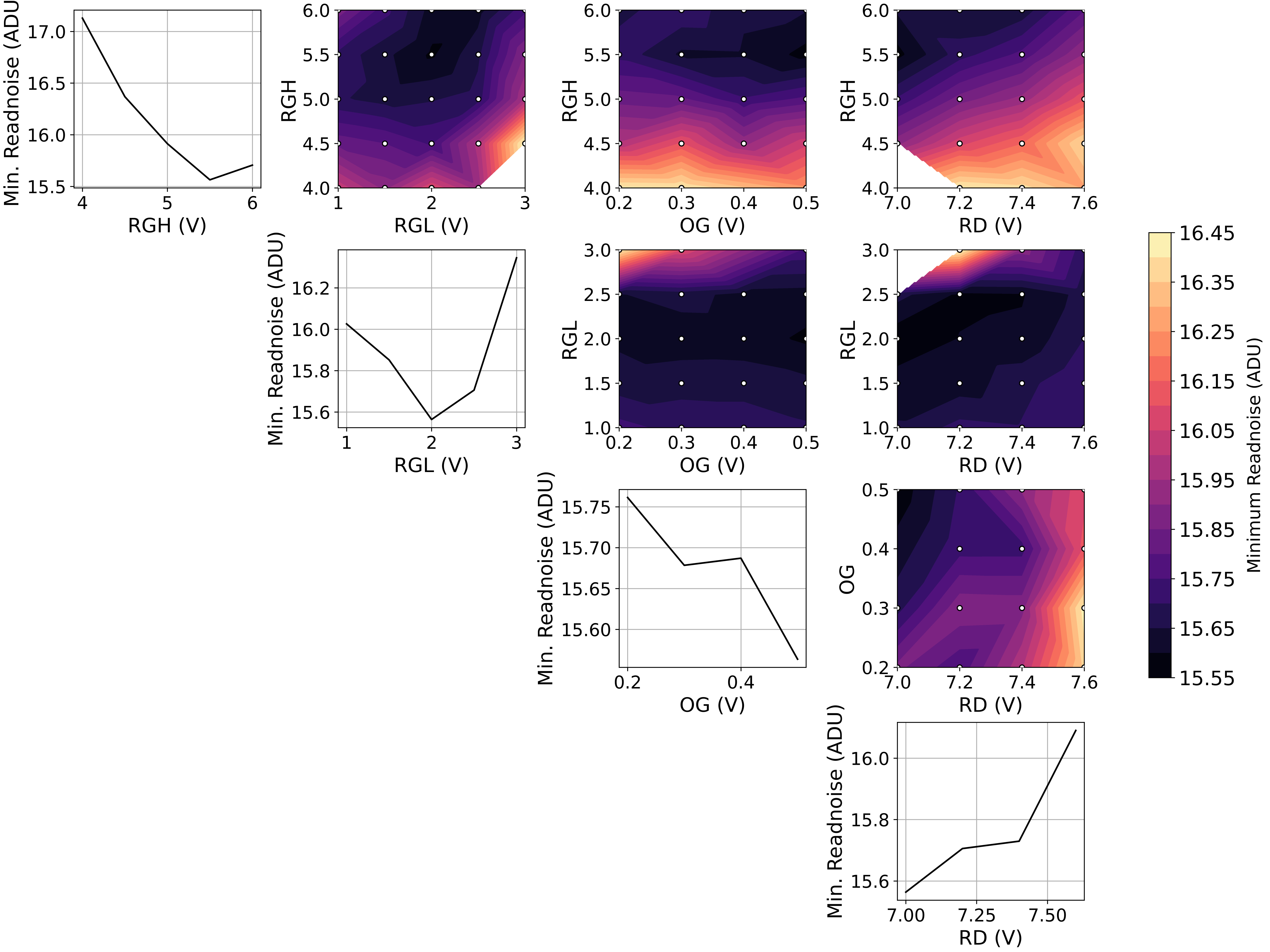}
   \end{tabular}
   \end{center}
   \caption[example] 
   { \label{fig:-100_triangle} 
Summary plots from a scan over RGH, RGL, OG, and RD. In the diagonal plots, the minimum read noise (in ADU) is reported at each value of RGH, RGL, OG, and RD that was scanned over. The off-diagonal plot elements show two-dimensional, linearly interpolated contour plots of the minimum read noise recorded for each combination of the four scan parameters. The white points on each contour plot represent parameter values where read noise values were actually measured. Gaps in the contours indicate parameter values for which the read noise was unphysically small, and did not yield spectral results.
}
\end{figure*}

\noindent Finding an optimal bias point for CCD detectors can be challenging, as the relationship between the various bias parameters and detector noise and gain can be complex. The performance of the detector is sensitive to the Reset Gate high- and low- state voltages (RGH and RGL, respectively), the Output Gate (OG) voltage, and the Reset Drain (RD) voltage. The locations of each of these bias nodes in the output stage of the CCID-93 detectors is indicated in Figure \ref{fig:output_stage}. The optimal combination of biases can also vary for different detectors, different test systems and readout electronics, and different temperatures. We describe here an approach for performing scans of a grid of values for RGH, RGL, OG, and RD that can be used to efficiently find an optimal operating point. 

Our scanning algorithm takes start, stop, and step voltage values for each of the bias parameters as inputs. It steps through each combination of RGH, RGL, OG, and RD values in the four-dimensional grid, taking five frames of data at each point. To estimate the read noise, we compute the standard deviation of the 50-column overscan region in each frame, then take the average across the five frames. Because the overscan consists of overclocked pixels with negligible thermal leakage, the measured variation reflects only the system's read noise. Assuming that the relationship between ADU and electron noise is relatively well behaved over the range of parameters under test, the algorithm returns the minimum read noise value in ADU as well as the corresponding bias voltages of the four scanned parameters. The triangle plot in Figure \ref{fig:-100_triangle} summarizes the results of a scan performed at 173\,K for the CCID-93 detector. Each scan measures read noise values for four hundred combinations of parameter values, taking a total of around 45 minutes to complete. These scans were kept relatively coarse to demonstrate the proof of concept, but in principle the approach could be used to scan over a more finely spaced grid. The diagonal plots in Figure \ref{fig:-100_triangle} track the optimal read noise over the range of values given for RGH, RGL, OG, and RD, where in each plot the recorded values are the minima at the given bias parameter for any combination of the other three parameters. The off-diagonal two-dimensional contour plots trace the minimum read noise over a grid of any combination of two given parameters, in order to visualize whether there may be a simply varying relationship or degeneracy between two sets of parameters, or whether a local or global minimum in the read noise has been found in that space.  

%The scanning algorithm takes start, stop, and step voltage values for each of the bias parameters as inputs. It steps over each combination of RGH, RGL, OG, and RD values on the four-dimensional grid of the parameter space, takes five frames of data, and calculates the read noise by taking the standard deviation in the overscan region of the average frame. The algorithm returns the minimum read noise value in analog digital units (ADU) as well as the corresponding bias voltages of the four scanned parameters. The triangle plot in Figure \ref{fig:-100_triangle} summarizes the results of a scan performed at 173\,K for the CCID-93 detector. Each scan presented here measures read noise values for four hundred combinations of parameter values, taking a total of around 45 minutes to complete. These scans were coarse to provide a proof of concept but in principle the algorithm can be used to scan over a more finely spaced grid. The diagonal plots in Figure \ref{fig:-100_triangle} track the read noise in the range of values given for RGH, RGL, OG, and RD, where in each plot the recorded noise values are the minima at the given bias parameter value for any combination of the other three parameters. The off-diagonal two-dimensional contour plots also trace the minimum read noise over a grid of any combination of two given parameters, in order to visualize whether there may be a simply varying relationship or degeneracy between two sets of parameters, or whether a local or global minimum in the read noise has been found in that space.  
 
While the noise of CCD detectors is expected to increase with temperature as the amount of leakage current increases, in some CCID-93 detectors the noise at warmer temperatures is higher than the expected trend. One possible reason could be that the bias point of the detector has some temperature dependence. To test whether the bias point of the detector drifts with temperature, we perform scans at three additional temperatures: 203\,K, 223\,K, and 243\,K. The summary triangle plots of read noise versus bias voltage for the three additional temperatures are presented in the Appendix \ref{sec:appendix}, while a summary of the optimal RGH, RGL, RD, and OG voltages are reported in Table \ref{tab:bias_params}. We compare the noise performance of the detector at 243\,K, 223\,K, 203\,K, and 173\,K for a default set of parameters, given as RGH=5.0\,V, RGL=2.0\,V, OG=0.4\,V, and RD=7.3\,V, to the optimal sets found at each temperature using the scanning method. A summary is provided in Table \ref{tab:param_scan}. 

The purpose of this procedure is to quickly and straightforwardly converge on an ideal operating point for any given detector. Ultimately, this scanning algorithm will be used to efficiently configure the bias for each of the 16 output channels of the AXIS prototype CCID-100 detectors at the anticipated focal plane temperature of 163\,K, optimizing the noise performance at the frame rates targeted by the AXIS mission.

\section{Results and Discussion}
\label{sec:results}

\begin{figure*}[ht!]
   \begin{center}
   \begin{tabular}{c}
   \includegraphics[width=0.8\textwidth]{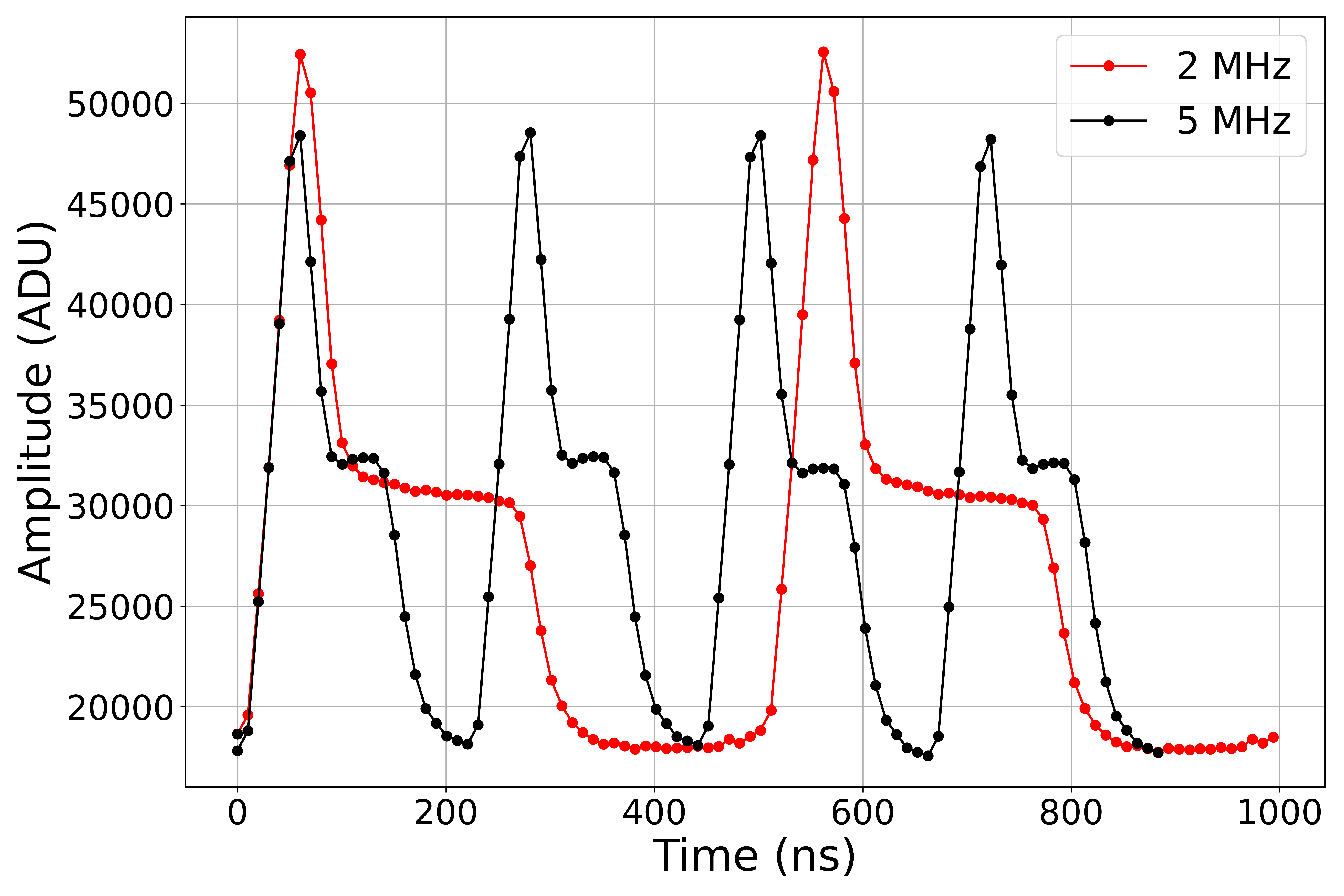}
   \end{tabular}
   \end{center}
   \caption[example] 
   { \label{fig:2MHz_5MHz_WF} 
Comparison of 2 MPixel/s with 5 MPixel/s readout rate waveform from the CCID-93 detector. Such high readout rates have been enabled by our development of onboard fast SW and RG clocks.  
}
\end{figure*}

\begin{figure*}[ht!]
   \begin{center}
   \begin{tabular}{c}
   \includegraphics[width=0.65\textwidth]{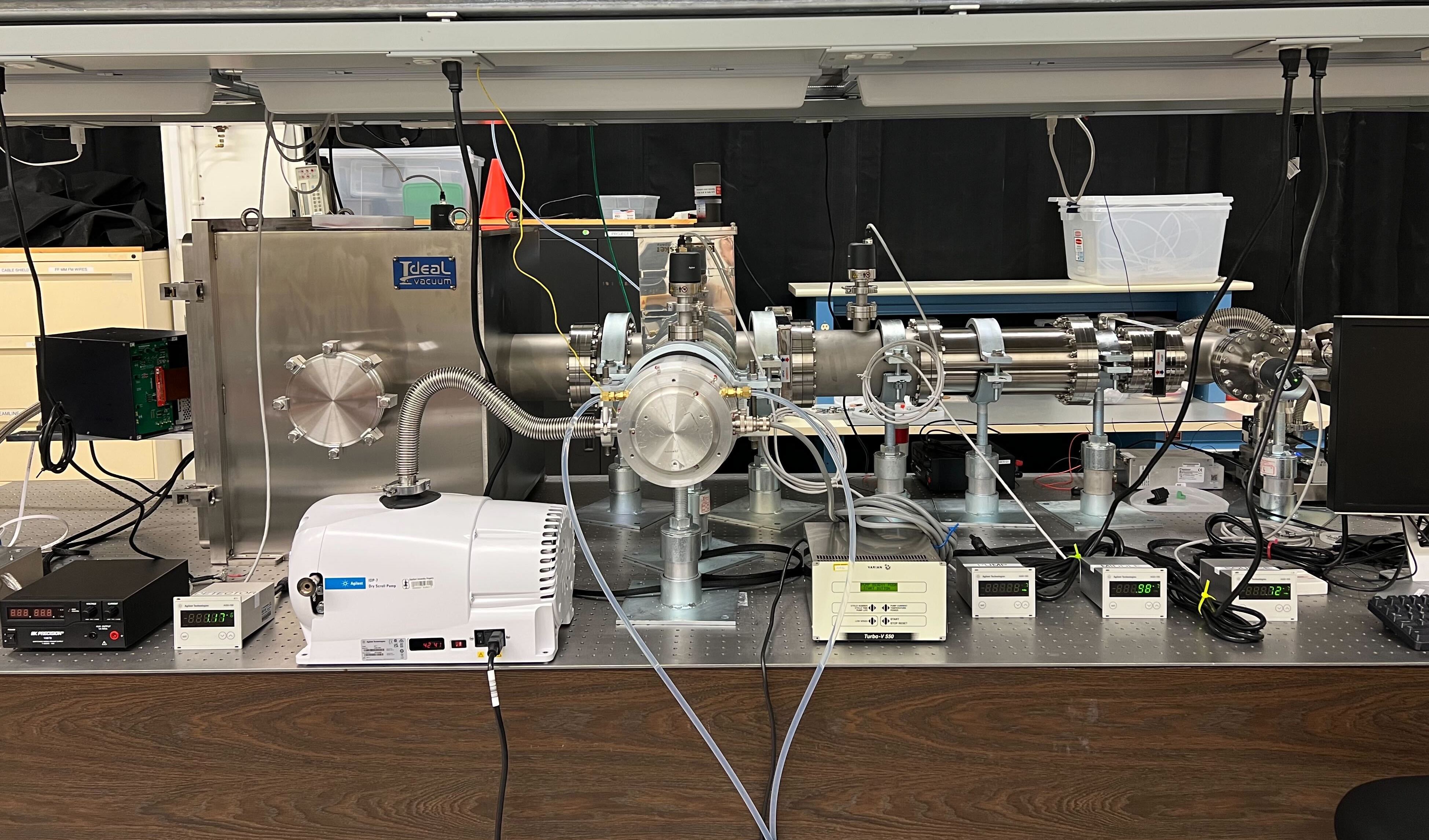}
   \includegraphics[width=0.29\textwidth]{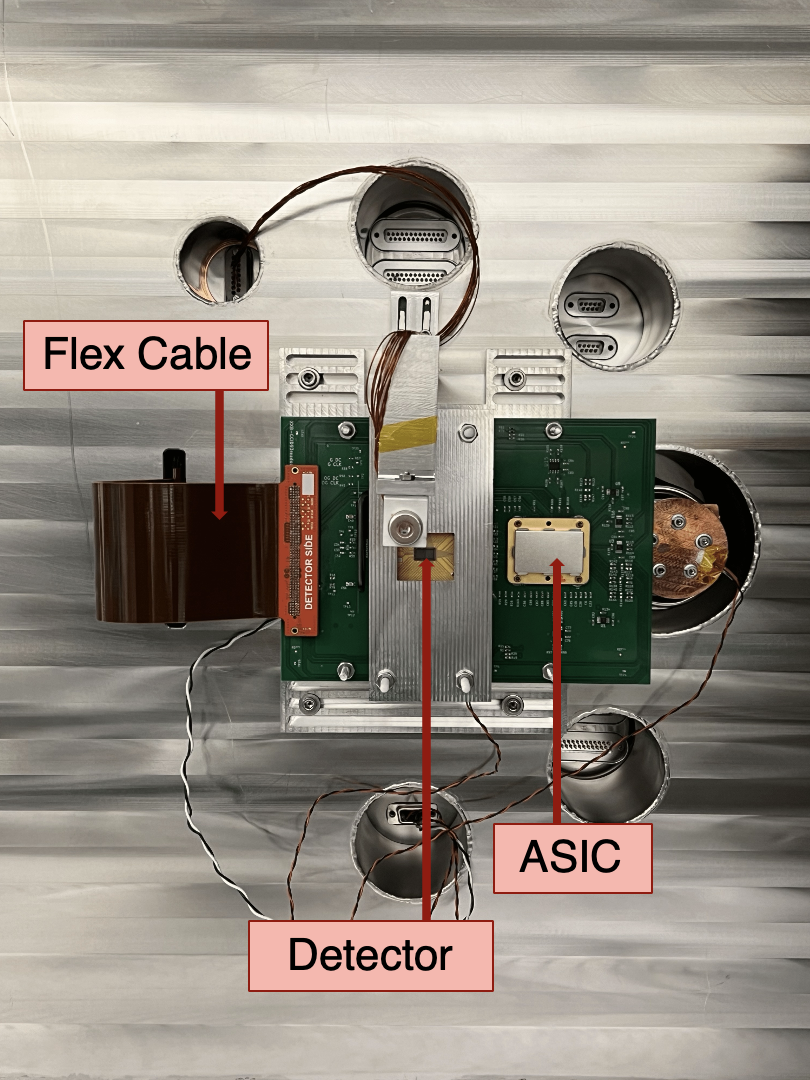}
   \end{tabular}
   \end{center}
   \caption[example] 
   { \label{fig:beamline} 
{\it Left:} Side view of the Gen 1.0 XOC X-ray Beamline CCD test chamber. {\it Right:} CCID-93 detector mounted inside the beamline with the MCRC ASIC and vacuum potted flex cable for transfer of signals between the detector (in vacuum) and the Archon (in atmosphere). Details of the experimental setup can be found in [\citenum{10.1117/12.3017691}].
}
\end{figure*}

\begin{figure*}[ht!]
   \begin{center}
   \begin{tabular}{c}
   \includegraphics[width=0.8\textwidth]{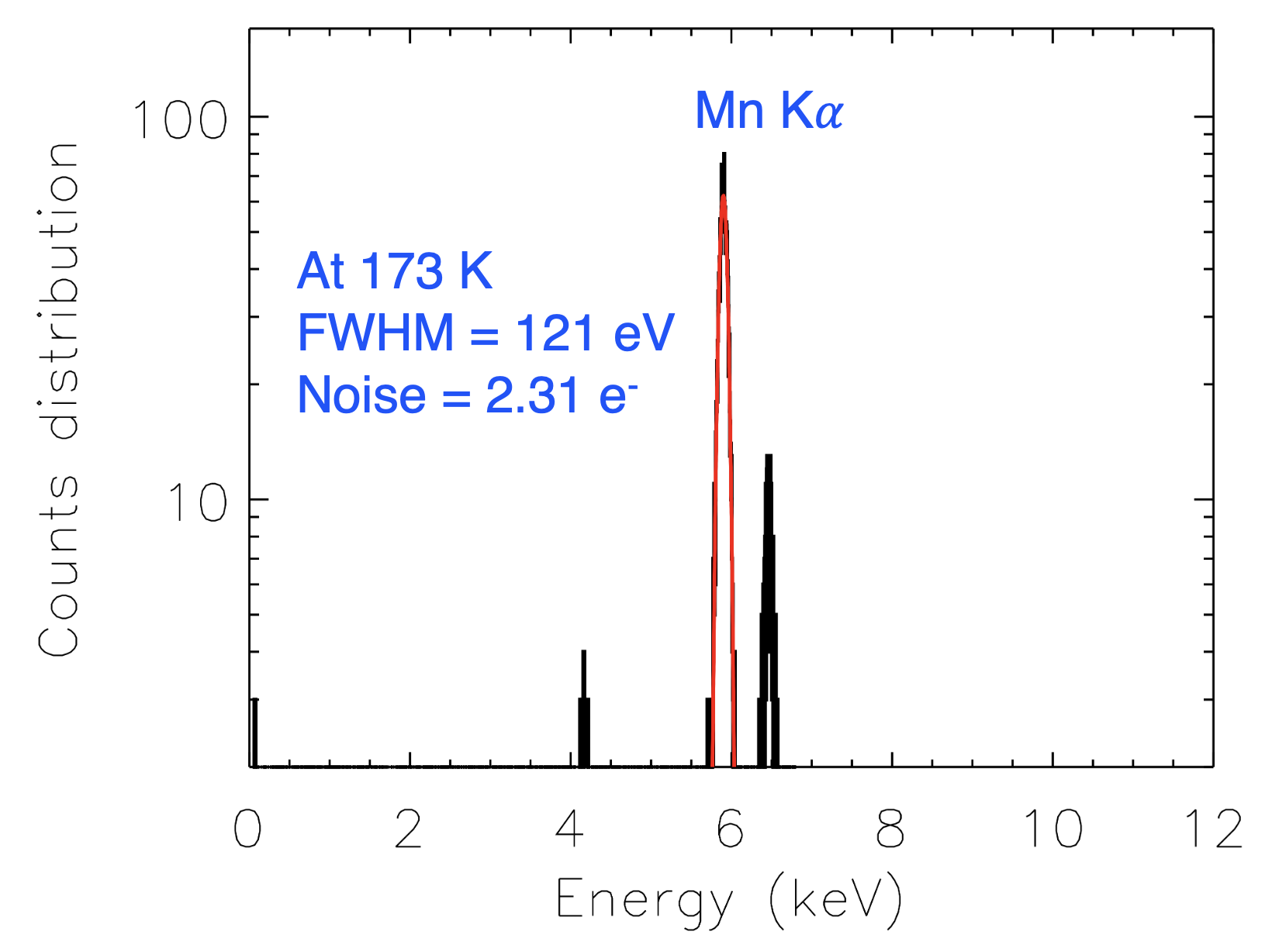}
   \end{tabular}
   \end{center}
   \caption[example] 
   { \label{fig:-100_spectrum} 
Spectrum obtained for an Fe-55 radioactive source at 173\,K from the CCID-93 detector with ASIC readout at 2 Mpixels/s serial transfer speed. This spectrum was taken following the implementation of the fast SW and RG clocking modes and the optimal bias parameters obtained from the 173\,K parameter scan were used (see Table \ref{tab:bias_params}). Overall noise is $\rm 2.31\, e^-$, with a full width half maximum (FWHM) of 121\,eV at 5.9\,keV.
}
\end{figure*}

\noindent Combining improvements to clock speeds from the implementation of onboard SW and RG clocks outlined in Section \ref{sec:SWRG}, which reduced noise at all speeds and enabled up to 5 MPixels/s readout (see Fig. 
\ref{fig:2MHz_5MHz_WF}), with the parameter scanning optimization methods described in Section \ref{sec:param_scan}, we present a spectrum obtained from an Fe-55 radioactive source using an MIT-LL CCID-93 and MCRC ASIC readout. The detector was cooled to 173\,K in the Gen 1.0 XOC X-ray Beamline vacuum test chamber (Fig. \ref{fig:beamline}, see [\citenum{10.1117/12.3017691}] for full details on the beamline test system) and read out at a serial transfer rate of 2 MPixels/s. The spectrum is shown in Figure \ref{fig:-100_spectrum}. At 5.9\,keV, the measured full width half maximum (FWHM) of the Mn $\rm K\alpha$ line was 121 eV, with a measured read noise of $\rm 2.31\,e^-$. 
%The noise levels and FWHM we measure at the readout speed of 2\,MPixels/s ($\rm \sim 7\,frames/s$ for the CCID-93) at 173\,K surpass the threshold requirements of the AXIS mission of 2\,MPixels/s serial rate, 5\,frames/s frame rate, and $\rm 3\,e^-$ read noise\cite{millerAXIS2023}. 

Table \ref{tab:bias_params} summarizes the optimal bias points found at each temperature over which we scanned the RGH, RGL, OG, and RD biases, as described in Section \ref{sec:param_scan}. We find that the preferred bias voltages for RG, OG, and RD vary with temperature. While it was found that different combinations of OG and RD could yield similar noise performance, the optimal RGH and RGL were sensitive to the temperature of the detector. RGH appears to prefer higher positive voltages at higher temperatures, while RGL prefers lower positive voltages at higher temperatures. The exact physical mechanism driving this trend remains under investigation. The noise measured using the default set of parameters versus the noise measured using the optimal set of parameters at each temperature at a readout speed of 2 MPixels/s is summarized in Table \ref{tab:param_scan}. Using the optimal operating point, the detector noise is decreased by up to 18\% at 243\,K, from $ \rm 5.24\,e^- $  to $ \rm 4.29\,e^- $. At this temperature, the optimal bias point was furthest from the default bias. The results suggest that optimizing the bias of these detectors at their intended operating temperature will be vital to achieving best performance, as will be finding the optimal bias in each output for multi-channel detectors such as the AXIS CCID-100. 

\begin{table}[H]
    \caption{Optimized bias settings (RGH, RGL, OG, RD) found at different temperatures for the CCID-93 detector using the methods described in Section \ref{sec:param_scan}.}
    \label{tab:bias_params}
    \begin{center}
    \begin{tabular}{|c|c|c|c|c|}
    \hline
    \textbf{Temp (K)} & \textbf{RGH (V)} & \textbf{RGL (V)} & \textbf{OG (V)} & \textbf{RD (V)} \\
    \hline
    243 & 6.0 & 1.0 & 0.2 & 7.6 \\
    223 & 6.0 & 1.5 & 0.3 & 7.2 \\
    203 & 5.5 & 1.5 & 0.5 & 7.0 \\
    173 & 5.5 & 2.0 & 0.5 & 7.0 \\
    \hline
    \end{tabular}
    \end{center}
\end{table}

\begin{table}[H]
    \caption{Best read noise achieved with the CCID-93 detector and ASIC readout at serial transfer speeds ranging from 2 Mpixels/s to 5 MPixels/s at 173\,K using the onboard RG and SW clocks described in Section \ref{sec:SWRG}. All noise values were achieved prior to performing a bias parameter optimization scan as described in Section \ref{sec:param_scan}.}
    \label{tab:SWRG_noise}
    \begin{center}
    \begin{tabular}{|c|c|c|}
    \hline
    \textbf{Serial Speed (MHz)} & {\textbf{Noise (e$^-$)}} & {\textbf{FWHM (eV) at 5.9\,keV}} \\
    \hline
    2 & $2.39\pm0.01$ & $118.5\pm1.7$ \\
    3 & $2.93\pm0.02$ & $124.4\pm1.7$ \\
    4 & $3.26\pm0.02$ & $126.7\pm1.9$ \\
    5 & $3.84\pm0.02$ & $125.8\pm2.2$ \\
    \hline
    \end{tabular}
    \end{center}
\end{table}

\begin{table}[H]
    \caption{Read noise and energy resolution of the CCID-93 detector at various temperatures between 243\,K and 173\,K using the default bias settings (RGH=5.0\,V, RGL=2.0\,V, OG=0.4\,V, RD=7.3\,V) and the optimal bias settings found using the methods described in Section \ref{sec:param_scan} (see Table \ref{tab:bias_params}). All noise values reported for 2 MPixels/s serial transfer speed.}
    \label{tab:param_scan}
    \begin{center}
    \begin{tabular}{|c|cc|cc|}
    \hline
    \textbf{Temp (K)} & \multicolumn{2}{c|}{\textbf{Noise (e$^-$)}} & \multicolumn{2}{c|}{\textbf{FWHM (eV) at 5.9\,keV}} \\
    \cline{2-5}
     & Default Bias & Optimized Bias & Default Bias & Optimized Bias \\
    \hline
    243 & $5.24\pm0.04$ & $4.29\pm0.02$ & $132.1\pm1.9$ & $122.8\pm1.8$ \\
    223 & $4.46\pm0.03$ & $3.92\pm0.03$ & $121.9\pm1.9$ & $119.7\pm1.9$ \\
    203 & $3.85\pm0.02$ & $3.66\pm0.03$ & $120.0\pm1.8$ & $119.5\pm1.7$ \\
    173 & $2.39\pm0.01$ & $2.31\pm0.01$ & $118.5\pm1.7$ & $120.8\pm1.9$ \\
    \hline
    \end{tabular}
    \end{center}
\end{table}

% \clearpage
\section{Summary}
\label{sec:conclusion}

The XOC group at Stanford University with collaborators at MIT-LL and MKI are working towards enabling the frame rates and noise performance requirements of next generation X-ray imaging observatories. The combination of the MIT-LL single poly CCID-93 CCD with our dedicated MCRC V1.0 readout ASIC, and the integration of fast clock drivers (specifically for SW and RG) close to the CCD, enables order of magnitude higher readout speeds than legacy X-ray CCDs. In addition, we have developed an automatized optimization process for CCD bias voltages (RG, OG and RD) that finds the best operating points (and therefore noise performance) efficiently. These two approaches combined allow us to reliably deliver high frame rates with excellent readout noise. The results presented here demonstrate clearly that the level of performance required of the AXIS mission can be achieved with the MIT-LL CCD fabrication process and state-of-the-art electronics. 
%Though it has fewer total pixels than the AXIS CCID-100 prototype, the 512\,x\,512 pixel CCID-93 detector has just one output channel. The 1440\,x\,1440 pixel CCID-100 detector will have 16 readout channels and two 8-channel ASIC readout chips to read them in parallel.
We are now focusing efforts to characterize and optimize the CCID-100 and each of its 16 output channels, a process streamlined by the methods presented here. Similar procedures should be applicable to any future strategic X-ray mission requiring fast, low noise, large-format imaging detectors.

\section{Acknowledgments}

We acknowledge support from NASA APRA grant 80NSSC22K1921 and NASA SAT grant 80NNSC23K0211, as well as support from NASA for the AXIS Probe Phase A study, under contract 80GSFC25CA019.

\clearpage
\begin{appendix} 

\section{Bias Read Noise Scan Triangle Summary Plots}
\label{sec:appendix}

\begin{figure*}[ht!]
   \begin{center}
   \begin{tabular}{c}
   \includegraphics[width=0.95\textwidth]{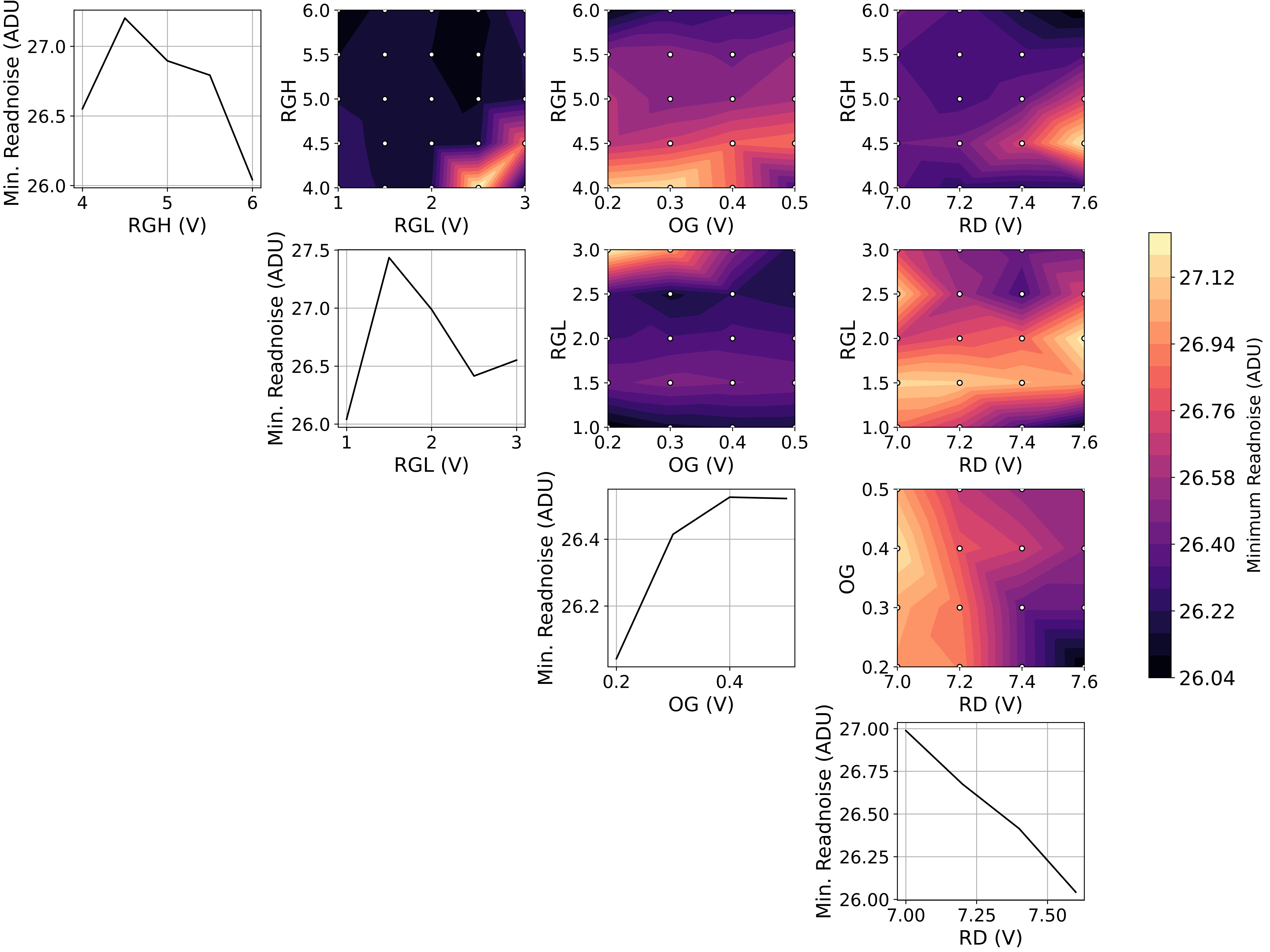}
   \end{tabular}
   \end{center}
   \caption[example] 
   { \label{fig:-30_triangle} 
Same RGH, RGL, OG, RD parameter scan summary plots as in Figure \ref{fig:-100_triangle}, but performed for a CCID-93 detector at 243\,K.
}
\end{figure*}

\begin{figure*}[ht!]
   \begin{center}
   \begin{tabular}{c}
   \includegraphics[width=0.95\textwidth]{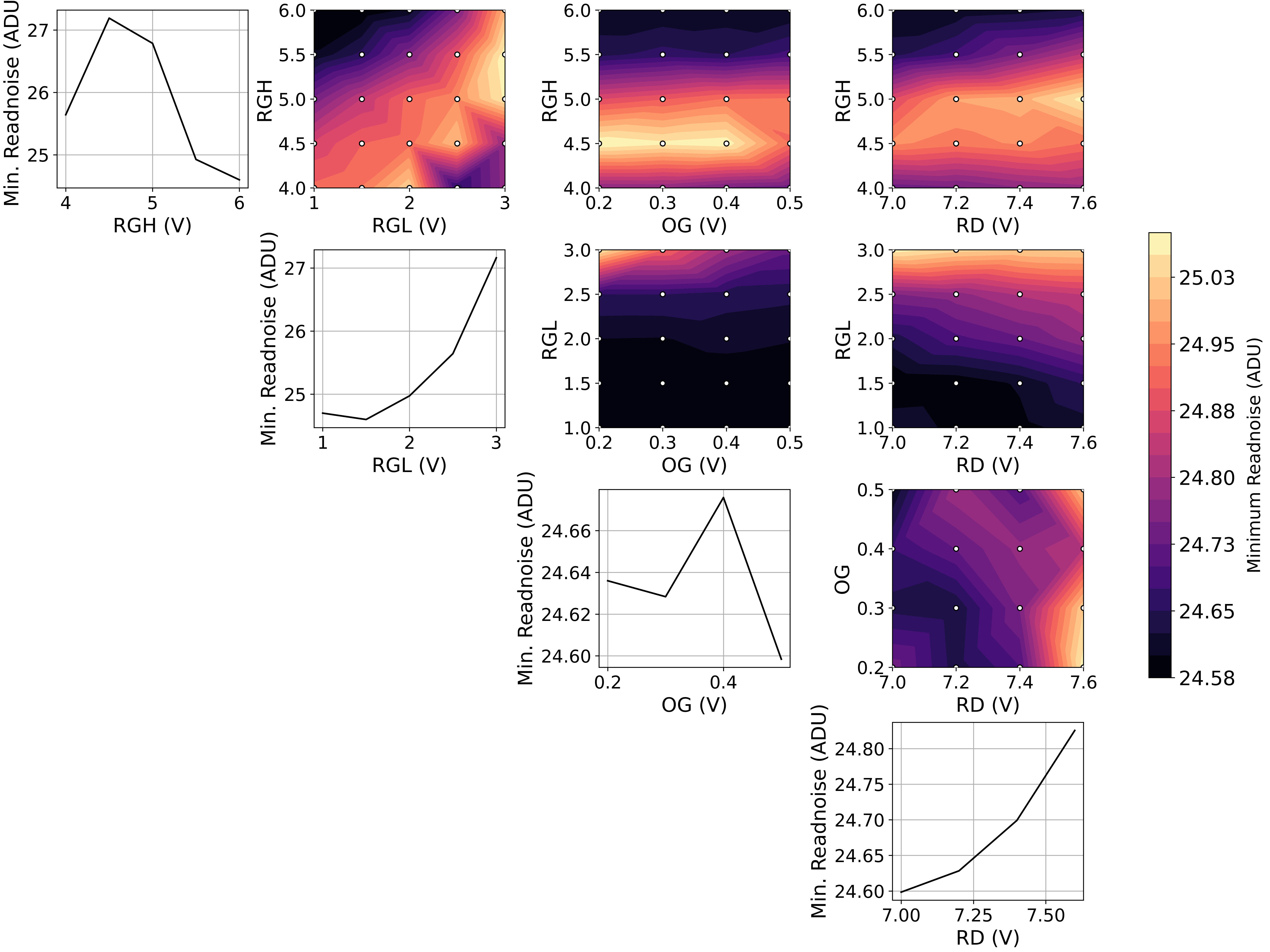}
   \end{tabular}
   \end{center}
   \caption[example] 
   { \label{fig:-50_triangle} 
Same RGH, RGL, OG, RD parameter scan summary plots as in Figure \ref{fig:-100_triangle}, but performed for a CCID-93 detector at 223\,K.
}
\end{figure*}

\begin{figure*}[ht!]
   \begin{center}
   \begin{tabular}{c}
   \includegraphics[width=0.95\textwidth]{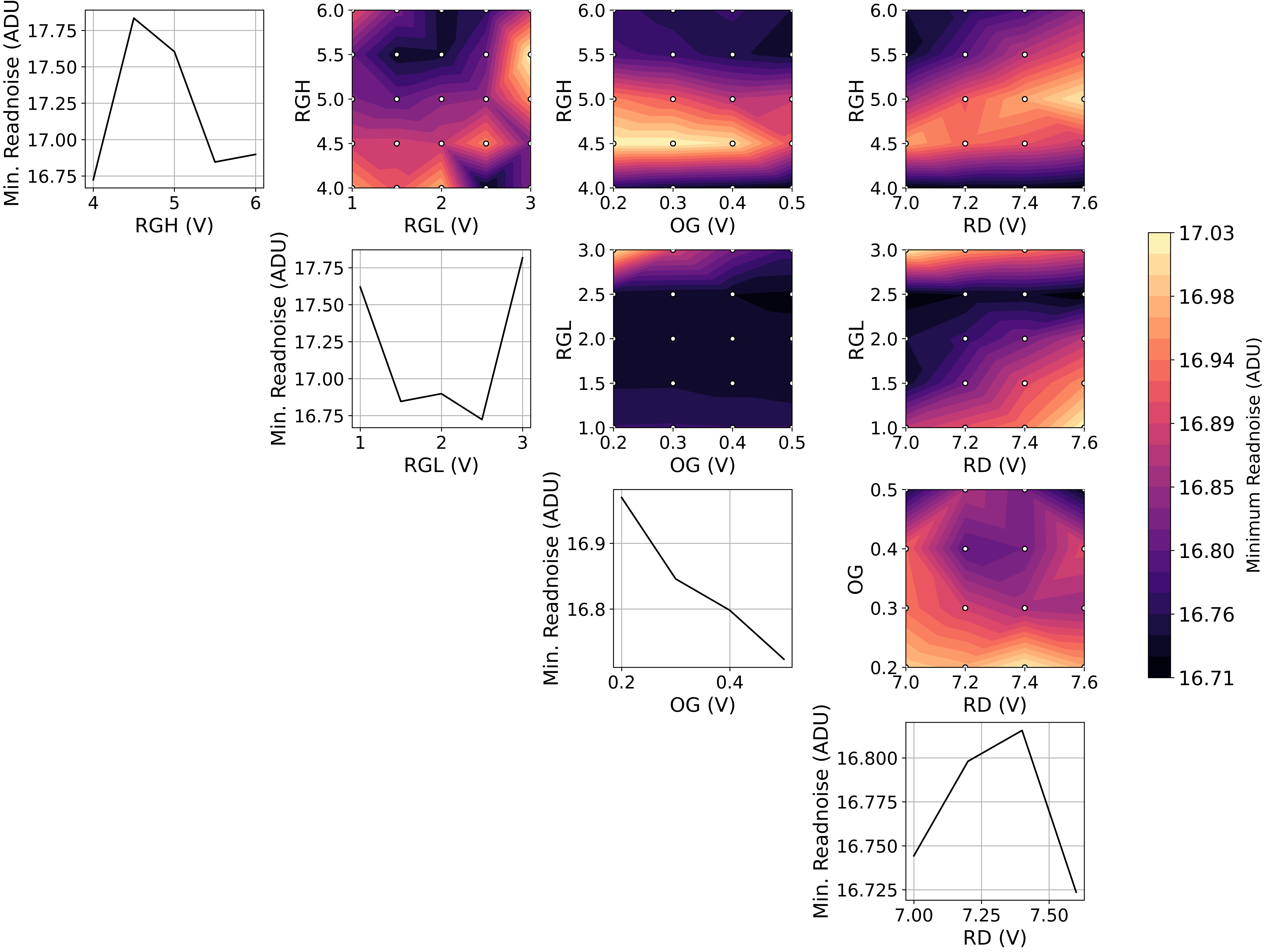}
   \end{tabular}
   \end{center}
   \caption[example] 
   { \label{fig:-70_triangle} 
Same RGH, RGL, OG, RD parameter scan summary plots as in Figure \ref{fig:-100_triangle}, but performed for a CCID-93 detector at 203\,K.
}
\end{figure*}

\end{appendix}

\clearpage
% References
\bibliography{report} % bibliography data in report.bib
\bibliographystyle{spiebib} % makes bibtex use spiebib.bst

\end{document}